\pdfoutput=1
\documentclass{Interspeech2024}

\usepackage[caption=false,font=normalsize,labelfont=sf,textfont=sf]{subfig}
\usepackage{textcomp}
\usepackage{stfloats}

\usepackage{xparse}
\usepackage{amsmath}
\usepackage{commath}
\usepackage{amssymb}
\usepackage{mathtools}
\usepackage{bbm}
\usepackage{physics}
\usepackage{siunitx}
\usepackage{trfsigns}
\usepackage{pifont}
\usepackage{xspace}
\usepackage{etoolbox}
\usepackage{multirow}
\usepackage{csquotes}
\usepackage{graphicx}
\usepackage[nosort]{cite}   %

\usepackage{tablefootnote}

\usepackage{hyperref}
\usepackage[capitalize]{cleveref}
\usepackage [table]{xcolor}
\usepackage{balance} %
\usepackage{algpseudocode}
\usepackage{algorithm}
\usepackage{booktabs}

\usepackage{url}

\newcolumntype{H}{>{\setbox0=\hbox\bgroup}c<{\egroup}@{}}   %

\newcommand{\mrow}[2][2]{\multirow{#1}[2]{*}{\begin{tabular}{@{}c@{}}#2\end{tabular}}}  %
\NewDocumentCommand{\tab}{ O{c} O{c} m }{\begin{tabular}[#2]{@{}#1@{}}#3\end{tabular}}   %

\newcommand*{\thl}{\fontseries{b}\selectfont}
\robustify\thl
\robustify\fontseries

\robustify\underline

\usepackage[normalem]{ulem}
\newcommand{\diff}[2]{\if\relax\detokenize{#1}\relax\else\textcolor{red}{\sout{\textcolor{black}{#1}}}\fi\textcolor{green}{#2}}
\newcommand{\changed}[1]{}

\usepackage{ulem}
\colorlet{lightgreen}{green!30}
\colorlet{lightred}{red!30}
\definecolor{forestgreen(web)}{rgb}{0.13, 0.55, 0.13}

\makeatletter
\newcommand\hladd{%
  \bgroup
  \markoverwith{\textcolor{lightgreen}{\rule[-.5ex]{.1pt}{2.5ex}}}%
  \ULon}
\newcommand\hlsout{%
  \bgroup
  \markoverwith{\textcolor{lightred}{\rule[-.5ex]{.1pt}{2.5ex}}\llap{\rule[.5ex]{.3pt}{0.4pt}}}%
  \ULon}
\makeatother
\DeclareRobustCommand{\diff}[2]{\hlsout{#1}\hladd{#2}}

\DeclareRobustCommand{\diff}[2]{#2}

\newcommand{\vect}[1]{\ensuremath{\boldsymbol{\mathbf{#1}}}}

\newcommand{\T}{^\mathrm{T}}

\usepackage{pgfplots} %
\usepgfplotslibrary{groupplots}
\usepackage{tikzsymbols}
\usepackage{pgfplotstable}

\newlength\fheight %
\newlength\fwidth %
\newlength{\figureheight} %
\newlength{\figurewidth} %

\usepackage{tikz}
\pgfplotsset{compat=1.9}

\usetikzlibrary{external}

\usetikzlibrary{arrows}
\usetikzlibrary{patterns}
\usetikzlibrary{backgrounds}
\usetikzlibrary{fit}
\usetikzlibrary{positioning}
\usetikzlibrary{shapes.geometric}
\usetikzlibrary{calc}
\usetikzlibrary{shapes.multipart}
\usetikzlibrary{shapes.misc}
\usetikzlibrary{shapes.symbols}
\usetikzlibrary{patterns.meta}
\usetikzlibrary{fadings}

\pgfdeclarelayer{bg}    %
\pgfsetlayers{bg,main}  %

\tikzset{>=stealth}

\tikzstyle{block}=[
    draw,
    text depth=0pt,
    thick, 
    rectangle,
    text centered,
    minimum height=3.4ex,
    rounded corners=0.3em,
    fill=black!6,
    inner sep=0.6em,
    ] %

\tikzstyle{dnn}=[]
\tikzstyle{enhBlock}=[]%
\tikzstyle{estBlock}=[dashed]%

\tikzstyle{branch}=[{circle,inner sep=0pt,minimum size=0.3em,fill=black}]
\tikzstyle{box}=[rectangle, rounded corners, draw=black, line width=1pt, text width=2cm]

\tikzstyle{arrow}=[{}-{>}, thick]
\tikzstyle{line}=[thick]
\tikzstyle{reverse arrow}=[{<}-{}, thick]

\tikzset{%
	do path picture/.style={%
		path picture={%
			\pgfpointdiff{\pgfpointanchor{path picture bounding box}{south west}}%
			{\pgfpointanchor{path picture bounding box}{north east}}%
			\pgfgetlastxy\x\y%
			\tikzset{x=\x/2,y=\y/2}%
			#1
		}
	},
	sin wave/.style={do path picture={    
			\draw [line cap=round] (-3/4,0)
			sin (-3/8,1/2) cos (0,0) sin (3/8,-1/2) cos (3/4,0);
	}},
	cross/.style={draw, circle, do path picture={    
			\draw [line cap=round] (-2/5,-2/5) -- (2/5,2/5) (-2/5,2/5) -- (2/5,-2/5);
	}},
	plus/.style={draw, circle, do path picture={    
			\draw [line cap=round] (-3/5,0) -- (3/5,0) (0,-3/5) -- (0,3/5);
	}},
	mic/.style={inner sep=0pt, do path picture={
			\draw (0,0) circle (0.9);
			\draw [line cap=round] (-0.9, -0.9) -- (-0.9, 0.9);
	}},
	mux/.style={trapezium, draw}
}

\usepackage[acronym,shortcuts]{glossaries}

\glsdisablehyper
\newacronym{SDR}{SDR}{Signal-to-Distortion Ratio}

\newacronym{CSS}{CSS}{Continuous Speech Separation}
\newacronym{GSS}{GSS}{Guided Source Separation}
\newacronym{PIT}{PIT}{Permutation Invariant Training}
\newacronym{uPIT}{uPIT}{utterance-level Permutation Invariant Training}
\newacronym{MSE}{MSE}{Mean Squared Error}
\newacronym{DFS}{DFS}{Depth First Search}
\newacronym[]{BLSTM}{BLSTM}{Bidirectional Long-Short-Term Memory}
\newacronym{DPRNN}{DPRNN}{Dual-Path Recurrent Neural Network}
\newacronym{WER}{WER}{Word Error Rate}
\newacronym{DER}{DER}{Diarization Error Rate}
\newacronym{ASR}{ASR}{Automatic Speech Recognition}
\newacronym{VAD}{VAD}{Voice Activity Detection}
\newacronym{vMFMM}{vMFMM}{von-Mises-Fischer Mixture Model}
\newacronym{NN}{NN}{Neural Network}
\newacronym{STFT}{STFT}{Short-Time Fourier Transform}

\newacronym{cACGMM}{cACGMM}{complex Angular Central Gaussian Mixture Model}

\newacronym{SLR}{SLR}{Segment-Level Speaker Reassignment}

\newacronym{EEND}{EEND}{End-to-End Neural Diarization}
\newacronym{NSD}{NSD}{Neural Speaker Diarization}

\newacronym{SMM}{SMM}{Spatial Mixture Model}

\newacronym{cpWER}{cpWER}{concatenated minimum-permutation Word Error Rate}

\newacronym{SC}{SC}{Spectral Clustering}

\newacronym{TSVAD}{TS-VAD}{Target-Speaker Voice Activity Detection}
\newacronym{TSSEP}{TS-SEP}{Target-Speaker Separation}

\interspeechcameraready

\title{Once more Diarization: Improving meeting transcription systems through segment-level speaker reassignment}

\name[]{Christoph}{Boeddeker}
\name[]{Tobias}{Cord-Landwehr}
\name[]{Reinhold}{Haeb-Umbach}

\address{Paderborn University, Germany}
\email{\{boeddeker,cord,haeb\}@upb.de}


\begin{document}

\maketitle

\begin{abstract}
Diarization is a crucial component in meeting transcription systems to ease the challenges of speech enhancement and attribute the transcriptions to the correct speaker.
Particularly in the presence of overlapping or noisy speech, these systems have problems reliably assigning the correct speaker labels, leading to a significant amount of speaker confusion errors.
We propose to add segment-level speaker reassignment to address this issue.
By revisiting, after speech enhancement, the speaker attribution for each segment, speaker confusion errors from the initial diarization stage are significantly reduced.
Through experiments across different system configurations and datasets, we further demonstrate the effectiveness and applicability in various domains.
Our results show that segment-level speaker reassignment successfully rectifies at least 40\% of speaker confusion word errors, highlighting its potential for enhancing diarization accuracy in meeting transcription systems.

\end{abstract}

\section{Introduction}
Meeting transcription systems aim to provide accurate recollections of natural conversations by answering the questions \enquote{Who?}, \enquote{What?}, and \enquote{When?}.
These systems typically consist of components to perform diarization, signal enhancement, and \gls{ASR}. 
The best processing order of these three tasks is, however, unclear.
On the one hand, many signal enhancement approaches benefit from an accurate initial diarization stage (e.g.\ \gls{GSS} \cite{Boeddeker2018GSS} or SpeakerBeam \cite{Zmolikova2017SpeakerBeam, Delcroix2021MeetinSpeakerBeam}) to determine whether overlapping speech is present and which speakers are active in a segment.
At the same time, overlapping speech is a main error source of those ``diarization-first'' systems \cite{cord2024geodesic}, so performing diarization on already enhanced and separated audio signals is significantly easier \cite{von2023meeting}.

A traditional approach to diarization is based on clustering as described in \cite{tranter2006overview}. It consists of first cutting the arbitrarily long input into segments, followed by speaker embedding extraction from each segment, and then clustering the embeddings.
While many improvements of this generic architecture have been proposed over the years \cite{shum2011exploiting,sell2014speaker,garcia2017speaker,wang2018dvector}, one inherent design flaw is its inability to properly  handle overlapping speech, where more than one speaker is active at a time.
This issue has been alleviated to some degree \cite{raj2021multi,kwon2022multi,cord2024geodesic}, but not finally solved.
With \gls{EEND} \cite{fujita19eend}, Fujita et al.{} proposed an alternative that conducts diarization with a single \gls{NN}, that is naturally able to handle overlapping speech.
Here it turned out, that, while it works well in a local context, it is suffering on long recordings and has difficulties re-identifying a speaker who has been silent for some time \cite{horiguchi2020end}.
To compensate for these deficiencies, other, hybrid systems either combine a local \gls{EEND} system with clustering-based diarization \cite{bredin2020pyannote, kinoshita2022tight} or use clustering-based diarization to infer enrollment embeddings that can be used for neural \gls{TSVAD} \cite{Medennikov20TS-VAD} or personal VAD \cite{ding2019PersonalVAD}.

Many current meeting transcription systems use these hybrid diarization approaches as the first processing stage.
Especially on real recordings, like CHiME-6 \cite{Watanabe2020CHiME6} and AMI \cite{carletta2005ami}, ``diarization-first'' approaches are predominantly used because the training of diarization systems on real long-form audio is straightforward. 
On the contrary, the training of a source separation stage on real long-form recordings is usually impossible because of unavailability of the training targets, the separated signals of the individual speakers in the mixture.

However, diarization-first systems have also issues, as mentioned earlier.
While the boundaries of speech activity can be detected relatively accurately, the models are prone to confuse speakers. This is because the quality of the speaker embeddings computed from a segment can be poor.
In \cite{kinoshita2022tight} it was shown that the number of speaker confusions steadily grows with the number of active speakers, amounting to \SIrange{35}{60}{\percent} of the total diarization errors, or \SIrange{6}{17}{\percent} absolute for more than \num{4} active speakers.
Compared to the tasks of speaker verification or speaker identification, where systems achieve error rates of \SIrange{1}{4}{\percent} absolute, but for significantly higher numbers of possible speakers \cite{nagrani2020voxceleb}, this demonstrates a large potential for improvement.
Furthermore, when set into the context of  \gls{ASR}, speaker confusions have 
a high impact on the \gls{WER}.
While missed or falsely detected speech at most result in errors for a single speaker, speaker confusions result in word errors for two speakers: deletions for one and insertions for the other speaker, even for an otherwise perfect transcription.  

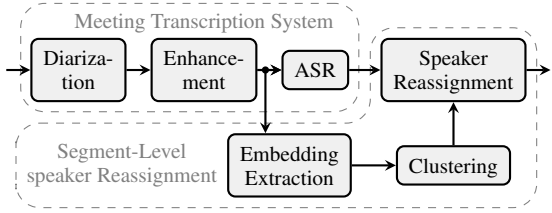
\begin{figure}[t]
  \centering
  {
    \begin{tikzpicture}[every node/.style={font=\footnotesize}]

    \tikzstyle{block}=[
        draw,
        text depth=0pt,
        thick, 
        rectangle,
        text centered,
        minimum height=3.4ex,
        rounded corners=0.3em,
        fill=black!6,
        inner sep=0.5em,
        ] %

    \node[block, align=center] (dia) at (0,0) {Diariza-\\tion};
    \node[block, anchor=west, align=center] (enh) at ($(dia.east) + (1em,0)$) {Enhance-\\ment};
    \node[block, anchor=west] (asr) at ($(enh.east) + (1em,0)$) {ASR};
    \node[block, anchor=west, align=center] (sr) at ($(asr.east) + (1.3em,0)$) {Speaker\\Reassignment};

    \node[text=gray, above, anchor=south, rotate=0, inner sep=0] (mtstext) at ($(enh.north)+(0,0.2em)$) {Meeting Transcription System};
    \node[fit=(dia)(enh)(asr)(mtstext), inner sep=0.4em, draw, dashed, gray, rounded corners=1em, shift={(0,0em)}] (mts) {};

    \coordinate (branchpos) at ($(enh.east)!1/3!(asr.west)$);
    
    \node[block, anchor=north, align=center] (emb) at ($(branchpos|-enh.south) + (1em,-1.3em)$) {Embedding\\Extraction};
    \node[block, align=center] (sc) at ($(emb-|sr) + (0em,0em)$) {Clustering};
    
    \node[text=gray, above, anchor=east, align=center, rotate=0, inner sep=0] (sldtext) at ($(emb.west)+(-0.4em,0)$) {Segment-Level\\speaker Reassignment};
    
    \node[fit=(emb)(sc)(sldtext), inner sep=0.4em, rounded corners=1em, shift={(0,0em)}] (fit1) {};
    \node[fit=(sr)(sc), inner sep=0.4em, rounded corners=1em, shift={(0,0em)}] (fit2) {};
    \draw[draw, draw, dashed, gray, rounded corners=1em] (fit1.north west) -| (fit2.north west) -- (fit2.north) -| (fit2.east) |- (fit1.south) -| (fit1.west) -- cycle; 

    \draw[arrow] (dia.west) +(-1em, 0) -- +(0,0);
    \draw[arrow] (dia) -- (enh);
    \draw[arrow] (enh) -- (asr);
    \draw[arrow] (enh) -- (asr);
    \draw[arrow] (asr) -- (sr);
    \draw[arrow] (branchpos) node[branch] {} -- (emb.north-|branchpos);
    \draw[arrow] (emb) -- (sc);
    \draw[arrow] (sc) -- (sr);
    
    \draw[arrow] (sr.east) --  +(1em, 0);

\end{tikzpicture}
  }
  
  \caption{
    Overview of a meeting transcription pipeline and its extension with the proposed \gls{SLR}. The audio segments after enhancement are input to the embedding extraction, and the \gls{ASR} output is then reassigned to match the new speaker identities.
  }
  \label{fig:overview}
\end{figure}

In this work, we propose an additional \gls{SLR}\footnote{\url{https://github.com/fgnt/speaker_reassignment}} component to address this issue. 
It extends the classical meeting transcription pipeline, as visualized in \cref{fig:overview}.
It takes the enhanced signals within the segment boundaries of the initial diarization stage and computes speaker embeddings from them.
Then, these embeddings are clustered to assign new speaker labels to each segment. Despite its simplicity, it results in a significant reduction of speaker assignment errors compared to the initial diarization. This is because the signals had been cleaned up and separated by the intermediate enhancement stage.
To increase the robustness of clustering, we also modify \gls{SC} to account for the noisiness of speaker embeddings extracted from short segments.
To underscore the efficacy of our proposed concept, we illustrate its versatility across diverse scenarios and systems.

\section{Segment-level speaker reassignment}
\label{sec:SLD}

The structure of the \acrfull{SLR} is based on the clustering-based diarization pipeline \cite{sell2014speaker,garcia2017speaker}, which
consists of a segmentation, an embedding extraction, and a clustering stage.
For the \gls{SLR}, the segments are directly obtained from the speech enhancement stage of a meeting transcription system as indicated in \Cref{fig:overview}.
Then, a speaker embedding extractor, e.g.\ a x-vector \cite{snyder2018xVector} or d-vector \cite{wang2018dvector} system, is used to extract a speaker embedding $\vect{e}_i$ for each segment $i$ in a meeting.
These embeddings are then clustered, e.g.\ with \gls{SC} or k-means, to group the segments that belong to the same speaker, and each segment is then assigned its new speaker label,
as illustrated in \cref{fig:illustration_sld}.

\subsection{Spectral Clustering}

Spectral clustering is widely used in clustering-based diarization systems.
It employs an adjacency matrix to derive new feature vectors, usually with reduced dimensionality compared to the original ones.
To achieve this, the process involves computing the (normalized) Laplacian matrix, whose eigenvectors are then used to derive the new features.
Subsequently, another clustering algorithm is applied to these new features.
While k-means could be used, in this work the algorithm from \cite{Yu2003SC}, also referred to as \enquote{discretize} in sklearn \cite{scikit-learn}, is utilized.

\subsubsection{Adjacency matrix}

The adjacency or similarity matrix $\vect{A} \in \mathbb{R}^{S\times S}$ with the entries $A_{i, j}$ for $i,j \in \{1, \dots, S\}$ consists of similarity scores (here: cosine similarities) between the $i$'th and $j$'th segment
\begin{align}
    A_{i,j} &= \begin{cases} \frac{\left|\vect{e}_i\T \vect{e}_j\right|}{\norm{\vect{e}_i} \norm{\vect{e}_{\smash{j}\vphantom{i}}}} & \text{if } i \neq j \\
    0 & \text{otherwise}
    \end{cases}
\end{align}
where $\vect{e}_i$ is an embedding for the $i$'th segment.
Note, that the diagonal entries are zero and not the self-similarity (i.e.\ one).

\begin{figure}[t]
  \centering
  {
    \input{figure/sld}
  }
  
  \caption{
    \acrfull{SLR}
  }
  \label{fig:illustration_sld}
\end{figure}

\subsubsection{Normalized Laplacian matrix}
To obtain the normalized graph Laplacian matrix we use the normalization proposed in \cite{shi2000normalized}.
The diagonal entries of the degree matrix $D_{i, i}$ are the sum of the row or column of the adjacency matrix
\begin{align}
    D_{i, i} &= \sum\limits_{m} A_{m, i} = \sum\limits_{m} A_{i, m}
\end{align}
The entries $L_{i, j}$ of the normalized Laplacian matrix $\vect{L}$ can then be calculated with
\begin{align}
    L_{i, j} &= \begin{cases}
        -\frac{A_{i, j}}{
            \sqrt{D_{i, i}}
            \sqrt{D_{j, j}}
        } & \text{if } i \neq j \\
        1 & \text{otherwise}
    \end{cases} %
\end{align}

\subsubsection{Feature transformation \& clustering}
The  $H$ eigenvectors, that belong to the $H$ smallest eigenvalues, $\vect{v}_{h} \in \mathbb{R}^{S}$, $h \in \{1,\ldots , H\}$, with the entries $v_{h,i}$ are calculated from the normalized Laplacian matrix $\vect{L}$.
Those entries $v_{h,i}$ of the eigenvectors are stacked and interpreted as features 
\begin{align}
    \vect{f}_i = \left[v_{1, i}, \dots, v_{H, i} \right]\T \in \mathbb{R}^{H}
\end{align}
for the $i$'th datapoint, i.e.\ each eigenvector contributes only one value to each feature vector $\vect{f}_i$.
Next, as mentioned earlier, the \enquote{discretize} clustering is used to obtain $K$ clusters.
We used the default number of features from sklearn: $H=K$.

\subsection{Attenuated Adjacency matrix}
\label{sec:attenuationAdjacencyMatrix}
A known issue of speaker embeddings is that their quality degrades with decreasing segment size, from which the embedding is computed \cite{zhou2021resnext}.
To account for this, we propose to attenuate entries in the adjacency matrix that stem from short segments 
\begin{align}
    \tilde{A}_{i,j} &= A_{i,j} \cdot c_{i, j}.
\end{align}
Here, entries of the adjacency matrix $A_{i,j}$ are attenuated by a factor $c_{i,j}$ that is computed based on the duration of the longer one of the two audio segments $i$ and $j$.
In this way, short segments cannot exhibit high similarities to each other, while still allowing for high similarities to long audio segments of the same speaker.
This prevents the accidental forming of clusters that only consist of short segments and enforces higher connectivity to long segments, thus accounting for the reliability of the speaker embeddings. 

Here, we consider two approaches, a step-wise attenuation 
\begin{align}
    c_{i, j} &= \begin{cases}
        1    & \text{if } 8 \le \max(T_i, T_j) \\
        \alpha^1    & \text{if } 4 \le \max(T_i, T_j) < 8 \\
        \alpha^2  & \text{if } 2 \le \max(T_i, T_j) < 4 \\
        \alpha^3    & \text{if } 1 \le \max(T_i, T_j) < 2 \\
        \alpha^4    & \text{if } \phantom{0 \le} \max(T_i, T_j) < 1, \\
    \end{cases} \label{eq:attenuation:stepwise}
\end{align}
where $0 \le \alpha \le 1$,
and a polynomial attenuation 
\begin{align}
    c_{i,j} &= \begin{cases}
        \left(\frac{\max(T_i, T_j)}{8}\right)^\beta & \text{if } \max(T_i, T_j) \le 8 \\
        1    & \text{otherwise},
    \end{cases} \label{eq:attenuation:polynomial}
\end{align}
where $\beta \ge 0$ and $T_i$ is the duration in seconds of a segment.

\section{Experiments}
\label{sec:experiments}

To demonstrate the effectiveness of the \gls{SLR} as postprocessing for meeting transcription pipelines, we evaluate it on different system configurations and datasets.
As datasets, we use  CHiME-6 \cite{Watanabe2020CHiME6},  DiPCo \cite{Segbroeck2019DiPCo}, and LibriCSS \cite{Chen2020LibriCSS}.
While CHiME-6 and DiPCo are considered challenging data sets, where current systems still exhibit high \glspl{WER}, LibriCSS is a comparatively easy data set, where the currently best \gls{WER} is at \SI{3.22}{\percent} \cite{taherian2023multi}.
For the meeting recognition systems, we choose pipelines with different diarization approaches (clustering-based diarization with and without \gls{EEND}, and TS-VAD-style diarization).

As a metric, we employ the \gls{cpWER}, which is computed as follows.
Transcriptions from the same speaker are concatenated for both the reference and the estimation.
As the permutation between reference and estimated transcription is unknown, the permutation that obtains the lowest \gls{WER} is used.
We use this metric instead of the \gls{DER} because the \gls{cpWER} also considers the content of the enhanced audio segments and considers speaker confusions for simultaneously active speakers as errors, contrary to the \gls{DER}. 
To have a lower bound for the \gls{SLR}, we calculate an oracle assignment, where the speaker label for each segment is chosen such that the \gls{cpWER} is minimal, similarly as in \cite{boeddeker2024tssep}.

\subsection{System configurations}

For CHiME-6 and DiPCo, we trained the CHiME-7 DASR baseline \cite{cornell2023chime}, a multi-channel system with \gls{EEND} coupled with clustering-based diarization, followed by \gls{GSS} and an ESPnet \gls{ASR} model.

On LibriCSS, three different model configurations were investigated.
The first system is a clustering-based diarization, where the embedding extractor is trained to yield frame-wise embeddings that allow to identify the active speakers even in regions of speech overlap. Those embeddings are  clustered using a \gls{vMFMM} \cite{cord2024geodesic}. Then, 
\gls{GSS} is applied on the diarization output, followed by the NeMo \gls{ASR} engine \cite{rekesh2023NemoASR}\footnote{\url{https://huggingface.co/nvidia/stt_en_fastconformer_transducer_xxlarge}}.
The last two configurations are a single-channel and multi-channel \gls{TSSEP} \cite{boeddeker2024tssep} system, an extension of \gls{TSVAD} that estimates speaker activity with time-frequency resolution instead of just time resolution.
For the multi-channel configuration of \gls{TSSEP}, \gls{GSS} is additionally applied
and source extraction is achieved by beamforming,
whereas the single-channel configuration estimates the enhanced signal by mask multiplication.
Both configurations use the same \gls{ASR} model from ESPnet \cite{chang2022WavLMASR}\footnote{\url{https://huggingface.co/espnet/simpleoier_librispeech_asr_train_asr_conformer7_wavlm_large_raw_en_bpe5000_sp}}.

Overall, the proposed \gls{SLR} is evaluated with a total of five different model configurations, consisting of three databases, three diarization, and two different speech enhancement approaches.

\begin{figure}[bt]
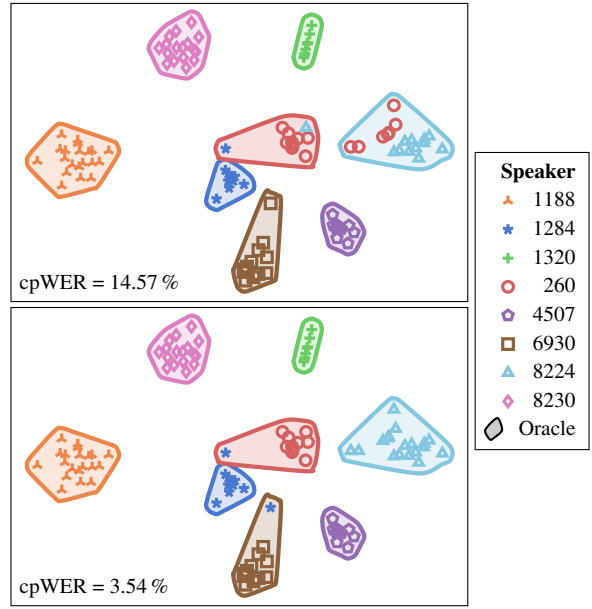

  \centering

    \begin{tabular}{l}
         {\providecommand{\old}{1}\input{figure/tsne}}  \\[-6.35em]
         {\providecommand{\old}{0}\input{figure/tsne}}
    \end{tabular}

  \caption{
    tSNE plot of the segment-level speaker embeddings for a LibriCSS example (session2, OV40) enhanced by \acrshort{TSSEP} before (upper figure) and after applying \acrfull{SLR} (lower figure).
    Many assignment errors have been fixed resulting in a decrease in \acrshort{cpWER} from \SI{14.57}{\percent} to \SI{3.54}{\percent}.
    The best possible assignment (oracle) results in a \acrshort{cpWER} of \SI{3.39}{\percent} on this session.
  }
  \label{fig:example_assignment}
\end{figure}

\subsection{Model details}

The speaker embeddings are extracted using a ResNet34-based d-vector system
from \cite{cord2023framewise}, 
trained on VoxCeleb \cite{nagrani2020voxceleb} augmented with MUSAN \cite{snyder2015musan} and simulated room impulse responses.
The same embedding extractor is applied to all investigated data sets without any fine-tuning on CHiME-6, DiPCo or LibriCSS.

We employ two different clustering techniques, k-means++ \cite{arthur2007kmeaspp} and \gls{SC}, the latter in its original form ($\alpha=1$, $\beta=0$) and with the modifications detailed in \cref{sec:attenuationAdjacencyMatrix}.
When employing k-means++ clustering, the embeddings are normalized to unit length. 
Both approaches (k-means++ and \gls{SC}) require the number of speakers as input parameter. It is taken from the preceding initial diarization stage.

In \cite{boeddeker2023chime7} a related \gls{SLR} method is proposed, to which we compare in the following. 
It relies on prototype availability and computes the distance between prototypes and embeddings acquired from the segments.
Instead of relying solely on distances, they employ a sticky approach, whereby a segment's label is altered only if the best distance surpasses all others by a margin.
In contrast, our method does not rely on prototypes and disregards the labels of the initial diarization stage.

\begin{table*}[th]
  \caption{\gls{cpWER} before and after \acrfull{SLR} for the different system configurations. Oracle denotes the best possible segment-level assignment that minimizes the \gls{cpWER}. Results marked with $^*$ are from a re-implementation of \cite{boeddeker2023chime7}.}
  \label{tab:results}
  \centering
  
\begin{tabular}{ l c c c c c c c }
    \toprule
    \mrow[2]{\bf Segment-\\\bf level\\\bf diarization} & \multicolumn{2}{c}{\textbf{Attenuation}} & \textbf{CHiME-6} & \textbf{DiPCo} & \multicolumn{3}{c}{\textbf{LibriCSS}} \\
    \cmidrule(lr){4-4} \cmidrule(lr){5-5} \cmidrule(lr){6-8}
    & $\alpha$ & $\beta$ & \tab{\bf CHiME-7 DASR\\\bf Baseline} & \tab{\bf CHiME-7 DASR\\\bf Baseline} & \tab{\bf vMFMM\\\bf + GSS} & \tab{\bf TS-SEP\\\bf + GSS} & \tab{\bf Single ch\\\bf TS-SEP} \\
    \midrule
    None & -- & -- & 62.25 & 58.16 & 12.62 & 5.36 &  7.81 \\
    \cite{boeddeker2023chime7} & -- & -- & 59.82 & 57.03 & 11.68\rlap{$^*$} &   5.08\rlap{$^*$} &   7.62\rlap{$^*$} \\
    \midrule
    k-means++ & -- & -- & 62.50 &  61.10 &  14.77& \bf 3.45 &   6.94 \\
    SC & \color{gray}$1$ & \color{gray}$0$ &  63.74 &  62.49 & 13.85 &   3.67 &   6.80 \\
    \midrule
     & $0.25$ & \color{gray}$0$  &  56.67 &  52.81 & 10.72 & 3.51 &   6.45 \\
    \smash{\tab[l]{SC: Step-wise\\attenuation}} 
    & $0.1$ & \color{gray}$0$ &  \bf 56.37 &  52.70 & 11.12 &   4.08 &   7.04 \\
    & $0$ & \color{gray}$0$ &  56.42 & \bf 52.49 &   15.36 &   6.27 &   8.91 \\
    \midrule
    & \color{gray}$1$ & $1$ &  57.80 &  55.17 & \bf 10.55 &   3.49 & \bf  6.42 \\
    & \color{gray}$1$ & $2$ &  57.67 &  53.77 & 10.66 &   3.48 &   6.51 \\
    \smash{\tab[l]{SC: Polynomial\\attenuation}}%
    & \color{gray}$1$ & $4$ &  56.71 &  53.11 & 10.83 &   3.50 &   6.43 \\
    & \color{gray}$1$ & $8$ &  56.39 &  52.82 & 12.13 &   3.88 &   6.75 \\
    & \color{gray}$1$ & $16$ & \bf 56.37 &  52.70 & 12.40 &   4.25 &   7.40 \\
    \midrule
    \color{gray} Oracle & -- & -- & \color{gray} 51.08 & \color{gray} 45.76 &   \color{gray}  \phantom{0}9.92 & \color{gray}  3.27 & \color{gray}   5.81 \\
    \bottomrule
\end{tabular}

\end{table*}

\subsection{Results}
\Cref{tab:results} summarizes the experimental results.
Here, it can be seen that especially for the model configurations that achieve already a good \gls{WER} (last two columns), \gls{SLR} is very effective (compare first results row with third and fourth).
\Cref{fig:example_assignment} shows an example of the speaker assignment before (upper part) and after (lower part) application of the \gls{SLR}.
The example is from a LibriCSS session, and the initial diarization has been obtained by multi-channel  \gls{TSSEP}.
One can clearly see that the cluster purity has improved.  
The reason for this improvement is the enhancement stage between the two diarization blocks. It leads to separated and cleaned-up signals, easing the task of the subsequent diarization.

The system configurations without \gls{TSSEP} do not profit from a simple clustering due to two main reasons: First, the higher oracle \glspl{cpWER} indicate a lower audio quality in the audio segments, which complicates the clustering.
Secondly, the models also output shorter audio segments. 
Especially due to the second point, ordinary clustering tends to combine multiple short embeddings that, due to their short length, are very noisy.
In this case, the speaker reassignment combines segments with allegedly high similarity, because the noisy embeddings pretend that the segments are similar.

To fix this second issue and to improve the robustness of the segment-level clustering, we used an attenuated adjacency matrix,
where the similarity value between short segments is decreased either with a step-wise (\cref{eq:attenuation:stepwise}) or a polynomial function (\cref{eq:attenuation:polynomial}), as described in \cref{sec:attenuationAdjacencyMatrix}.
Both have one hyperparameter and 
to distinguish between them, we use $\alpha \in [0, 1]$ and $\beta \geq 0$ for the step-wise and polynomial functions, respectively.
When $\alpha = 1$ or $\beta = 0$, the attenuation is one, and the approach reverts to plain spectral clustering.
With $\alpha = 0$, all similarities between short segments (of which the longer one is shorter than \SI{8}{\second}) are set to zero.

On CHiME-6 and DiPCo it is beneficial to attenuate similarities between short segments 
and even setting them to zero ($\alpha=0$) worked.
This can be explained, by the fact, that both have long recordings and each speaker has enough segments that are longer than 8 seconds, which is not the case for LibriCSS.
A less aggressive attenuation again significantly improves the results on LibriCSS for the \gls{vMFMM}, while also providing a benefit for \gls{TSSEP}.

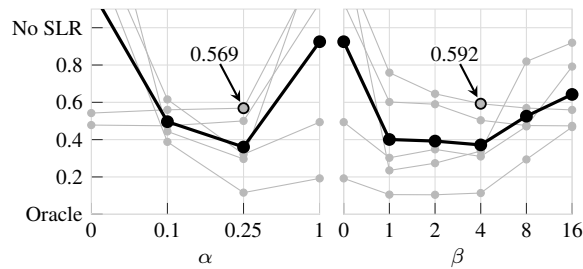
\begin{figure}[bt]
  \centering
  {\begin{tikzpicture}

\definecolor{darkslategray38}{RGB}{38,38,38}
\definecolor{gray}{RGB}{185,185,185}
\definecolor{lightgray204}{RGB}{220,220,220}

\pgfplotsset{
    ex_line_common_axis_opts/.style={
        gray, mark=*, mark size=0.15em,
    },
    mean_line_common_axis_opts/.style={
        very thick, black, mark=*, mark size=0.2em
    },
}

\begin{groupplot}[group style={group size=2 by 1,horizontal sep=1em}]

\nextgroupplot[
    axis line style={lightgray204},
    x grid style={lightgray204},
    xmajorgrids,
    xmajorticks=true,
    xmin=0, xmax=3,
    xtick={0,1,2,3},
    ytick pos=left,
    xtick pos=bottom,
    xticklabels={0,0.1,0.25,1},
    xminorgrids,
    xtick style={color=darkslategray38},
    ytick={0,0.2,0.4,0.6,0.8,1},
    yticklabels={Oracle,0.2,0.4,0.6,0.8,No \acrshort{SLR}},
    y grid style={lightgray204},
    ymajorgrids,
    ymajorticks=true,
    xlabel={$\alpha$},
    ymin=0, ymax=1.1,
    yminorgrids,
    ticklabel style={font=\footnotesize},
    label style={font=\footnotesize},
    ytick style={color=darkslategray38},
    scale only axis, %
    width=9.5em,
    height=8.6em,
    xtick=data,
]
\addplot [ex_line_common_axis_opts]
table[x expr=\coordindex] {%
0 0.478095527836933
0.1 0.47414055369638
0.25 0.500304228780043
1 1.13340432004868
};
\addplot[ex_line_common_axis_opts]
table[x expr=\coordindex] {%
0 0.542395693135935
0.1 0.559892328398385
0.25 0.568506056527591
1 1.34912516823688
};
\addplot[ex_line_common_axis_opts]
table[x expr=\coordindex] {%
0 2.01379839018781
0.1 0.443848217707934
0.25 0.296282100421617
1 1.45649674204676
};
\addplot[ex_line_common_axis_opts]
table[x expr=\coordindex] {%
0 1.43364403018198
0.1 0.387927208166889
0.25 0.115845539280959
1 0.192188193519751
};
\addplot[ex_line_common_axis_opts]
table[x expr=\coordindex] {%
0 1.54811715481172
0.1 0.615527661552766
0.25 0.321245932124593
1 0.493723849372385
};
\addplot[mean_line_common_axis_opts]
table[x expr=\coordindex] {%
0 1.20321015923088
0.1 0.496267193904471
0.25 0.360436771426961
1 0.92498765464489
};

\coordinate (tmp) at (axis cs:2,0.568506056527591);
\coordinate (offset) at (-1.2em,1.9em);
\draw[line width=0.1em,fill] (tmp) circle (0.2em);
\node[font=\footnotesize, above, inner sep = 0.05em, fill=white] (text) at ($(tmp)+(offset)$) {0.569};
\draw[arrow, shorten >=0.3em, shorten <=0.05em] (text) -- (tmp);

\nextgroupplot[
    axis line style={lightgray204},
    x grid style={lightgray204},
    xmajorgrids,
    xmajorticks=true,
    xmin=0, xmax=5,
    xminorgrids,
    xtick pos=bottom,
    xtick={0,1,2,3, 4, 5},
    xticklabels={0, 1, 2, 4, 8, 16},
    xtick style={color=darkslategray38},
    ytick={0,0.2,0.4,0.6,0.8,1},
    yticklabel=\empty,
    y grid style={lightgray204},
    ymajorgrids,
    ymajorticks=false,
    xlabel={\smash{$\beta$}\vphantom{$\alpha$}},
    ymin=0, ymax=1.1,
    yminorgrids,
    ytick style={color=darkslategray38},
    scale only axis, %
    ticklabel style={font=\footnotesize},
    label style={font=\footnotesize},
    width=9.5em,
    height=8.6em,
    xtick=data,
]
\addplot [ex_line_common_axis_opts]
table[x expr=\coordindex] {%
0 1.13340432004868
1 0.601916641314269
2 0.589747490112565
4 0.503802859750533
8 0.475509583206572
16 0.473836324916338
};
\addplot [ex_line_common_axis_opts]
table[x expr=\coordindex] {%
0 1.34912516823688
1 0.759084791386272
2 0.645760430686407
4 0.592462987886945
8 0.569313593539704
16 0.55935397039031
};
\addplot [ex_line_common_axis_opts]
table[x expr=\coordindex] {%
0 1.45649674204676
1 0.234955921809122
2 0.27366807205826
4 0.338827136834036
8 0.819087773093139
16 0.9195093905711
};
\addplot [ex_line_common_axis_opts]
table[x expr=\coordindex] {%
0 0.192188193519751
1 0.105193075898802
2 0.104305370616955
4 0.113626276076343
8 0.293830448291168
16 0.467376830892144
};
\addplot [ex_line_common_axis_opts]
table[x expr=\coordindex] {%
0 0.493723849372385
1 0.302649930264993
2 0.347745234774524
4 0.309623430962343
8 0.470943747094374
16 0.791724779172478
};
\addplot [mean_line_common_axis_opts]
table[x expr=\coordindex] {%
0 0.92498765464489
1 0.400760072134692
2 0.392245319649742
4 0.37166853830204
8 0.525737029044991
16 0.642360259188474
};

\coordinate (tmp) at (axis cs:3,0.592462987886945);
\draw[line width=0.1em,fill] (tmp) circle (0.2em);
\node[font=\footnotesize, above, inner sep = 0, fill=white, , inner sep = 0.05em] (text) at ($(tmp)+0.9*(offset)$){0.592};
\draw[arrow, shorten >=0.3em, shorten <=0.05em] (text)  -- (tmp);

\end{groupplot}

\end{tikzpicture}%
}
    \caption{
        Remaining relative transcription errors caused by speaker confusions w.r.t.\ \gls{cpWER} when using \gls{SLR}.
        \textit{No \acrshort{SLR}} denotes the starting performance,
        \textit{Oracle} the best possible assignment.
        Light gray lines show the performance of the five systems of \cref{tab:results}, the thick black line displays their mean.
    }
  \label{fig:attenuation_performance}    
\end{figure}

To get a better impression of the gain of the \gls{SLR}, the relative transcription error rate increase caused by
speaker confusions w.r.t.\ \gls{cpWER} is shown in \cref{fig:attenuation_performance} for the different system configurations and the mean across all systems.
We defined this error measure as follows: $0$ means the best possible assignment in terms of \gls{cpWER} is obtained, while $1$ means there was no improvement in \gls{cpWER} compared to skipping the \gls{SLR}. Any value in-between is the percentage of increase in \gls{cpWER} from the oracle result to the one without speaker reassignment.
Note that with $\alpha=0.25$ or $\beta=4$ it is possible to detect and fix at least \SI{40}{\percent} of segment-level speaker confusion independent of the scenario.

\section{Conclusions}
\label{sec:conclusion}
In this work, we proposed \acrfull{SLR} as a post-processing step for meeting transcription systems.
We also proposed a modification of the adjacency matrix of \gls{SC} to deemphasize the impact of noisy embedding vectors.
Our experiments show  a notable reduction in speaker confusion errors, with improvements of at least \SI{40}{\%} in \gls{cpWER} observed across a diverse set of experiments.
Notably, our proposed approach is effective for both high and low error-rate meeting transcription systems.
In one instance, on the LibriCSS dataset, our method reduced the WER from \SI{5.36}{\%} to \SI{3.45}{\%}, approaching the performance of the best possible speaker assignment, which is at \SI{3.27}{\%}.

\noindent\textbf{Open Source:}
The code for the \gls{SLR} is available on GitHub:\\
\resizebox*{\columnwidth}{!}{\url{https://github.com/fgnt/speaker_reassignment}}

\ifinterspeechfinal
    \section{Acknowledgements}
    
    Christoph Boeddeker was funded by Deutsche Forschungsgemeinschaft (DFG), project no. 448568305. 
    Computational resources were provided by BMBF/NHR/PC2.     
\else
\fi

\bibliographystyle{IEEEtran}
\bibliography{mybib}

\end{document}